\documentclass[aps,prb,reprint,superscriptaddress,twocolumn,longbibliography]{revtex4-2}

\usepackage{graphicx}
\usepackage{amsmath,amsfonts,amssymb,mathtools}
\usepackage{color}
\usepackage{subfigure}
\usepackage{physics}
\usepackage{dsfont}
\usepackage{hyperref}
\usepackage{booktabs}
\usepackage{url}
\usepackage{leftidx}
\usepackage{textcomp}
\usepackage{gensymb}
\usepackage{threeparttable}
\usepackage{rotating}
\usepackage{esvect}
\usepackage{booktabs}
\usepackage{marvosym}
\usepackage{lmodern}
\usepackage[noend]{algpseudocode}
\usepackage[dvipsnames]{xcolor}

\usepackage{dcolumn}
\usepackage{bm}
\usepackage{mathtools}
\usepackage{epstopdf}
\usepackage{esint}
\usepackage{xcolor}
\usepackage{nicefrac}
\usepackage{pifont}
\usepackage[normalem]{ulem}

\usepackage{svg}
\usepackage{hyperref}
\usepackage{cleveref}

\definecolor{emerald}{rgb}{0.31, 0.78, 0.47}
\definecolor{blue(ncs)}{rgb}{0.0, 0.53, 0.74}

\hypersetup{
     colorlinks=true,
     linkcolor=magenta,
     filecolor=blue,
     citecolor=emerald,      
     urlcolor =blue(ncs),
}

\usepackage{chngcntr}

\DeclareMathAlphabet{\pazocal}{OMS}{zplm}{m}{n}

\newcommand{\lsoc}{\lambda_{\rm soc}}

\newcommand{\Rv}{{\bf R}}

\newcommand{\up}{\uparrow}
\newcommand{\down}{\downarrow}
\newcommand{\la}{\langle} 
\newcommand{\ra}{\rangle} 

\newcommand{\hdelta}{\hat{\bm{\delta}}}
\newcommand{\hx}{\hat{\bf{x}}}
\newcommand{\hy}{\hat{\bf{y}}} 
\newcommand{\iv}{{\bf i}}
\newcommand{\jv}{{\bf j}}

\newcommand{\bea}{\begin{eqnarray}}
\newcommand{\eea}{\end{eqnarray}}

\newcommand{\noi}{\noindent}

\begin{document}

\title{Superconducting diodes from magnetization gradients}

\author{Mercè Roig}
\affiliation{Niels Bohr Institute, University of Copenhagen, DK-2200 Copenhagen, Denmark}

\author{Panagiotis Kotetes}
\affiliation{CAS Key Laboratory of Theoretical Physics, Institute of Theoretical Physics, Chinese Academy of Sciences, Beijing 100190, China}

\author{Brian M. Andersen}
\affiliation{Niels Bohr Institute, University of Copenhagen, DK-2200 Copenhagen, Denmark} 

\date{\today}

\vskip 1cm

\begin{abstract}
The superconducting diode effect may exist in bulk systems as well as in junctions when time-reversal and inversion symmetries are simultaneously broken. Magnetization gradients and textures satisfy both requirements and therefore also allow for superconducting diodes. We concretely demonstrate such possibilities in two-dimensional superconductors. We first consider superconducting Rashba metals in the presence of an inhomogeneous out-of-plane exchange field. Using analytical arguments, we reveal that such magnetization gradients stabilize a helical superconducting ground state, similar to homogeneous in-plane magnetic fields. Our predictions are confirmed by employing self-consistent real-space numerical lattice simulations exemplified through the cases of a uniform magnetization gradient or a ferromagnetic domain wall. Furthermore, by considering a phase difference, we determine the nonreciprocal current-phase relations and explore their parameter dependence. Our calculations show that planar devices with out-of-plane magnetization gradients may be as efficient supercurrent rectifiers as their analogs induced by uniform in-plane fields. In addition, they feature the advantage that by means of tailoring the spatial profile of the out-of-plane magnetization, one may optimize and spatially control the diode effect. Finally, we show that superconducting diodes may become also accessible even in the absence of spin-orbit coupling by means of suitable spatially-varying magnetization fields.
\end{abstract}

\maketitle

\section{Introduction} 

Recently, a series of expe\-ri\-ments have demonstrated the existence of nonreciprocal supercurrents in dif\-fe\-rent device setups, thus establishing the so-called supercon\-duc\-ting diode effect~\cite{Ando2020,Diez-Merida,Lyu2021,Shin21,Lin2022,Baumgartner2022,Strambini2022,Wu2022,Bauriedl2022,Hou22,Pal2022,Turini2022,Du2023,Gupta2023,Gutfreund2023,Narita2023,margineda2023sign}. The supercurrent rectification appears promising for future electronic logic circuit applications utilizing non-dissipative supercurrents and its efficiency depends strongly on the particular design of the diode. When the superconducting diode effect takes place, different critical currents  arise for current flow in opposite directions. This implies that there exist certain regimes of the current amplitudes, in which the junction is resistive along one direction and supercon\-duc\-tive along the other. Thus, the supercon\-duc\-ting diode effect can be equivalently viewed as a polarity-dependent metal-superconductor transition. This appealing rectification property can ori\-gi\-na\-te, for example, from vortex-related physics~\cite{Hou22,Gutfreund2023,margineda2023sign}, be a property of the tunnel-junction itself~\cite{Diez-Merida,Shin21,Strambini2022,Wu2022,Gupta2023}, or be rooted in finite-momentum Cooper pairing of the superconduc\-ting ground state~\cite{Lyu2021,Lin2022,Baumgartner2022,Bauriedl2022,Pal2022,Turini2022,Du2023,Narita2023}. In this work, we focus on the latter mechanism, i.e., supercurrent nonreci\-pro\-ci\-ty in Josephson tunnel junctions arising from helical superconducting ground states~\cite{Daido22,Daido_PRB22,Bergeret22,Yuan22,He_2022,Scammell_2022,Legg2022,Davydova22,Karabassov2022,ZhangPRX2022,kochan2023phenomenological,Costa2023}.

Obviously, the detailed mechanism and whether an external field is required for nonreciprocity or not, depends on the particular system under investigation. For the case of homogeneous noncentrosymmetric superconductors, the Rashba spin-orbit coupling (SOC) in conjunction with an in-plane magnetic field was shown to allow for nonreciprocity of the supercurrents~\cite{Daido22,He_2022,Bergeret22}. Essentially, in the presence of the field the ground state adiabatically transforms into the so-called the helical superconducting phase~\cite{Kaur2005,Agterberg2007}. In the latter, the Cooper pairs carry a finite momentum $\mathbf{q}_0$, whose direction is determined by the orientation of the external Zeeman field. As a consequence of the finite-momentum pairing, the critical depairing current naturally depends on the current direction compared to the ground state propagation direction of the Cooper pairs set by $\mathbf{q}_0$. 

The above mechanism is, however, not unique. For instance, it was recently discussed in a work that two of us co-authored~\cite{KotetesSuraAndersen}, that imposing an out-of-plane magnetization gradient on a planar superconductor with Rashba SOC can lead to additional ground state electrical currents. The emergence of the latter reflects the violation of time-reversal and inversion symmetries, and also forms the basis for a superconducting diode effect. Never\-the\-less, Ref.~\onlinecite{KotetesSuraAndersen} restricted the discussion to out\-li\-ning the possibility and the key concepts for such magnetization gradient diodes, without carrying out a detailed inve\-stigation of the arising Josephson diode effects. In fact, a number of interesting questions need to be answered in this respect. The first concerns whether an out-of-plane magnetization gradient also induces a helical pai\-ring state in a Rashba superconductor. Moreover, when it comes to possible applications, it is important to exa\-mi\-ne the efficiency of such diodes. Note that when con\-si\-de\-ring magnetization gradients, the exact spatial profile of the exchange field is crucial. Therefore, optimizing this feature constitutes an important study area for diodes using magnetization gradients.

In this work, we explore the superconducting diode effect induced from magnetization gradients in 2D systems, both in the presence and absence of Rashba SOC. Notably, the spatially varying magnetization can be either an intrinsic property of the system, e.g., spontaneously generated spin-density waves due to interactions, or, imposed by external means. We first investigate the emergence of a diode effect in a bulk Rashba superconductor under the influence of an out-of-plane magnetization gradient, and discuss the ari\-sing similarities and differences with its counterpart ge\-ne\-ra\-ted by uniform in-plane magnetization components. Afterwards, we extend our study to the Josephson diode effect, i.e., nonreciprocal Josephson transport arising in junction settings. In particular, we study 2D Josephson junctions composed of superconducting Rashba me\-tals, in which time-reversal and $C_2$ rotation symmetries become already violated at zero phase-bias due to the influen\-ce of the out-of-plane magnetization. We perform self-consistent real-space simulations of such Josephson junction devices to determine the real-space current di\-stribution, the asymmetric current-phase relation, and the associated directional-dependent critical currents. We indeed find that this setup features the Josephson diode effect and we study its parameter dependence in detail. Compared to the standard case with in-plane magnetic fields, we find that the nonreciprocal effect is equally pronounced for the magnetization gra\-dient junctions, and may be further optimized by proper designs of the spatial form of the magnetization gra\-dients.

We conclude this work by demonstrating that a spatially-varying magnetization profile is alone capable of generating Josephson diodes, without any requirement for SOC. This is possible when on top of the out-of-plane magnetization gra\-dient discussed earlier, an additional magnetization profile with a suitable structure is present. The latter profile can be, for instance, induced by a magnetic texture crystal whose magnetic moments exhibit a nontrivial winding in two spatial dimensions. Such infinitely repeating magnetic structures carry zero net magnetic moment since they preserve a non-symmorphic time-reversal symmetry~\cite{Ramazashvili,Kotetes_2013}. However, they ge\-ne\-ral\-ly lack inversion symmetry, hence effectively give rise to Rashba SOC~\cite{MostovoyFerro,Christensen_18,KotetesSPIE}. A number of works have theoretically predicted~\cite{BrauneckerSOC,KarstenNoSOC,Ivar,KlinovajaGraphene,Nakosai2013}, as well as experimentally~\cite{Kontos} demonstrated, the engineering of such a synthetic Rashba SOC. The latter SOC combined with the out-of-plane magnetization profiles discussed earlier, gives rise to the sought-after Josephson diode effect. Our analytical predictions are further backed by numerical investigations for a magnetic texture crystal of the spin whirl type~\cite{Christensen_18,KotetesSPIE,Huang2023}.

The paper is organized as follows: in Sec.~\ref{sec:phenomenologyhelical} we briefly review the phenomenology of helical superconductivity and magnetization-induced supercurrents, with a focus on the cases mediated by uniform in-plane fields and out-of-plane magnetization gradients. Section~\ref{sec:caseofrashba} considers the particular case of a Rashba metal and provides a detailed discussion of the amplitude of linear-response coefficients connecting the induced currents and the magnetization for that case.  In Sec.~\ref{latticebasics} we turn to the lattice model and the numerical setup for obtaining the current-phase relations in the Rashba metal. Section~\ref{sec:latticeresults} discusses the results from the lattice model, comparing the diode effects for different junction realizations. Next, in Sec.~\ref{sec:NoSOC}, we show using both analytical and numerical methods that the Josephson diode effect can emerge from a spatially-nonuniform magnetization even in the absence of Rashba SOC. Finally, section~\ref{sec:discussion} presents our discussion and conclusions.  

\section{Helical superconductivity in a Rashba metal}\label{sec:phenomenology}

Before proceeding with the detailed numerical inve\-sti\-ga\-tions of the superconducting diode effect in va\-rious experimentally relevant platforms, we first lay the foundations for the phenomena discussed here, and compare their mechanism to prior theoretical proposals. In the following two subsections, we first introduce a phenomenological description of the helical phase stabilized by (in)homogeneous magnetism and afterwards we focus on the case of a Rashba metal in the quasiclassical regime.

\subsection{Phenomenological theory of helical superconductivity in Rashba systems}\label{sec:phenomenologyhelical}

Our discussion can be facilitated by introducing the phenomenological energy density $E(\bm{r})$ of the Rashba superconductor, which is obtained after integrating out the electronic degrees of freedom. Here, $\bm{r}=(x,y)$ corresponds to a position in 2D space, with $x$ and $y$ the coordinates. Assuming a bulk system with a uniform superconducting gap, we focus on the couplings appearing among the superconducting phase field $\varphi(\bm{r})$, the in-plane components of the electromagnetic vector potential $A_{x,y}(\bm{r})$ and a magnetization/magnetic/exchange field $M_{x,y}(\bm{r})$, along with the gradients of the out-of-plane magnetization $\bm{\nabla}M_z(\bm{r})$ with $\bm{\nabla}=(\partial_x,\partial_y)$. After the analysis of Ref.~\onlinecite{KotetesSuraAndersen}, we find the following expression for $E(\bm{r})$:
\begin{align}
E(\bm{r})=\bm{{\cal A}}(\bm{r})\cdot\Big[D\bm{{\cal A}}(\bm{r})/2-\bm{J}_{\rm mag}(\bm{r})\Big],
\label{eq:EnergyDensity}
\end{align}

\noi where we introduced the gauge invariant vector potential $\bm{{\cal A}}(\bm{r})=\bm{A}(\bm{r})+\hbar\bm{\nabla}\varphi(\bm{r})/2e$ and the current induced by the in-plane magnetization and the out-of-plane-magnetization gradient:
\begin{align}
\bm{J}_{\rm mag}(\bm{r})=\Lambda\hat{\bm{z}}\times\bm{M}(\bm{r})+{\cal X}\bm{\nabla}\times\hat{\bm{z}}M_z(\bm{r})\,.
\label{eq:Jmag}
\end{align}

\noi We remark that throughout this work all vectorial quantities refer to the in-plane components, e.g., $\bm{M}(\bm{r})=(M_x(\bm{r}),M_y(\bm{r}))$. The only exception is the unit vector pointing in the $z$ direction which is denoted $\hat{\bm{z}}$. In the above, we introduced the reduced Planck constant $\hbar$, the electric charge unit $e>0$, the superfluid stiffness $D$, and the coefficients $\Lambda$ and ${\cal X}$ which control the strength of the magnetization-induced currents. 

The coefficient $\Lambda$ is responsible for the standard Edelstein effects~\cite{Edelstein1,Edelstein2}. While this coupling constant has been calculated in prior works~\cite{Pershoguba2015,KotetesSuraAndersen}, we briefly discuss it also here for completeness. The interconversion coefficient ${\cal X}$ has been investigated by two previous works for a superconducting Rashba metal. Firstly, Pershoguba {\it et al.}~\cite{Pershoguba2015} determined the behavior of this coefficient as a function of the ratio of the Fermi energy $E_F$ and the SOC energy at the Fermi level $E_{\rm soc}$, when the pairing gap $\Delta\geq0$ is considered negligible. Subsequently, Ref.~\onlinecite{KotetesSuraAndersen} investigated the limit in which the Fermi energy goes to infinity, and ${\cal X}$ is found as a function of the ratio of $E_{\rm soc}$ and $\Delta$. In addition to the contribution of the Rashba SOC, the latter work also considered the contributions of the Zeeman effect to ${\cal X}$, obtaining the following expressions:
\begin{align}
{\cal X}=\chi+\chi_Z\quad{\rm and}\quad \chi_Z=g\mu_B\chi_\perp^{\rm spin}/2\,.
\end{align}

\noi Notably, the coefficient ${\cal X}$ consists of two principal contributions $\chi_Z$ and $\chi$. The former stems from the Zeeman effect, while the latter incorporates the orbital and Rashba couplings to the magnetization. In the above, $g$ denotes the gyromagnetic Land\'e factor, $\mu_B$ the Bohr magneton, and $\chi_\perp^{\rm spin}$ defines the out-of-plane spin susceptibi\-li\-ty of the Rashba system.

We observe from Eqs.~(\ref{eq:EnergyDensity}) and (\ref{eq:Jmag}) that the free energy obtained for in-plane magnetic fields and out-of-plane magnetization gra\-dients has a similar form. Therefore, we expect the system to adiabatically transit to the helical phase with $\Delta(\bm{r})=\Delta {\rm exp}\big(i\mathbf{q}_0\cdot\bm{r}\big)$ in either scenario~\cite{Kaur2005,Agterberg2007}. The current of the system at zero vector potential is given by:
\begin{align}
\bm{J}(\bm{r})=-\left.\frac{\delta E(\bm{r})}{\delta \bm{A}(\bm{r})}\right|_{\bm{A}=\bm{0}}=-D\frac{\hbar}{2e}\bm{\nabla}\varphi(\bm{r})+\bm{J}_{\rm mag}(\bm{r})\,.
\end{align}

For a uniform $\bm{J}_{\rm mag}(\bm{r})=\bm{J}_{\rm mag}$, the system completely cancels out the magnetization-induced current. This is precisely the helical superconductivity regime, which is dictated by a zero total electric current. Hence, the resulting slope $|\mathbf{q}_0|$ of the superconducting phase profile in the helical phase is determined by minimizing the energy density of Eq.~\eqref{eq:EnergyDensity} with respect to the phase gra\-dient, and depends on the precise values of the coefficients $\Lambda$ and ${\cal X}$. The above implies that the system exhibits a nonzero constant superconducting phase gradient $\bm{\nabla}\varphi$ given by:
\begin{align}
\mathbf{q}_0=\bm{\nabla}\varphi=\frac{1}{D}\frac{2e}{\hbar}\bm{J}_{\rm mag}\,.
\label{eq:phasefromJmag}
\end{align}

\subsection{The case of a superconducting Rashba metal}\label{sec:caseofrashba}

At this point, it is important to discuss the parameter dependence of the coefficients $\Lambda$ and ${\cal X}$ in the representative case of a superconducting Rashba metal. In this case, the system is described in its normal phase by two helicity bands with energy dispersions 
\begin{align}
\varepsilon_\pm(p)=\frac{p^2-p_F^2\pm2m_*\upsilon p}{2m_*}\,,
\end{align}

\noi where $p=|\bm{p}|$ corresponds to the modulus of the momentum vector $\bm{p}=\hbar\bm{k}$, where $\bm{k}$ defines the wave vector.  Here, $p_F$ is the Fermi momentum in the absence of the Rashba SOC, whose strength is determined by the va\-riable $\upsilon>0$. The variable $m_*>0$ denotes the effective mass. For a typical metallic conductor considered here, which features an electron gas, $m_*$ can be approximated with the bare electron mass $m_e$ and the Land\'e factor with $g=2$. In the following, we consider the regime in which the Fermi energy $E_F=p_F^2/2m_e$ is much larger than $E_{\rm soc}=\upsilon p_F$ and $\Delta$. Thus, in this limit, $E_F$ can be taken to infinity. In addition, we restrict to the case $\delta=\Delta/E_{\rm soc}\ll1$.

Under the above conditions, Ref.~\onlinecite{KotetesSuraAndersen} has shown that $\chi$ reads as:
\begin{align}
\chi\big(0<\delta\ll1\big)=-\frac{e}{4\pi\hbar}\,,
\end{align}

\noi while the Zeeman contribution becomes:
\begin{align}
\chi_Z\big(0<\delta\ll1\big)=\mu_B\chi_\perp^{\rm spin}=\frac{e}{2\pi\hbar}\,.    
\end{align}

\noi Interestingly, the contribution of the Zeeman effect is substantial and in the present case is twice that of $\chi$, thus leading to $|{\cal X}|=\chi$. Since in the present case the sign of ${\cal X}$ is not crucial, we can take advantage of the coincidence of $|{\cal X}|$ and $\chi$, and do not explicitly take into account the Zeeman effect in our upcoming analysis. 

We now proceed with the expression of the coefficient $\Lambda$ for a superconducting Rashba metal in the quasiclassical limit and at zero tem\-pe\-ra\-tu\-re $T$. It is straightforward to show that:
\begin{align}
\Lambda=e\upsilon\chi_{||}^{\rm spin},
\end{align}

\begin{figure}[t!]
\begin{center}
\includegraphics[angle=0,width=.95\linewidth]{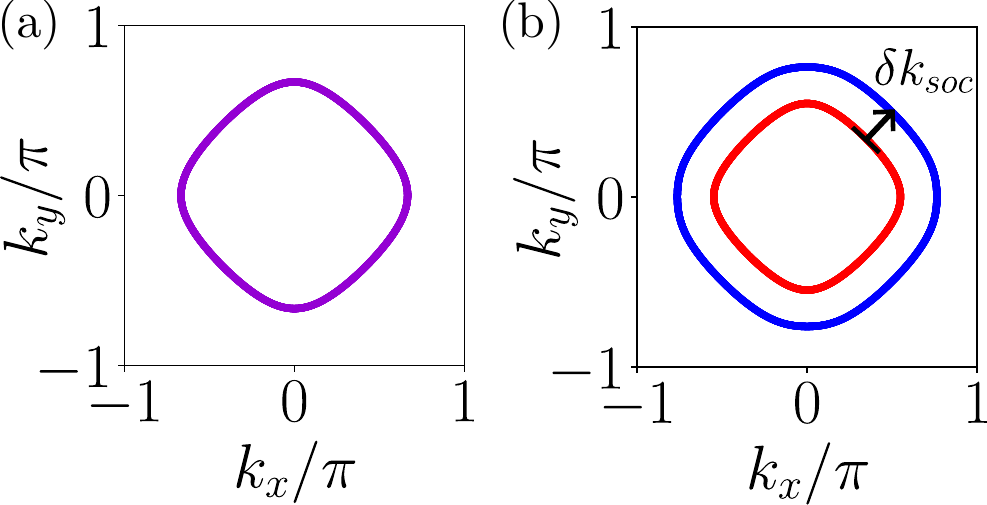}
\caption{Fermi surface for (a) a spin-degenerate metal, and (b) a metal with Rashba spin-orbit coupling, which features an inner (red) and an outer (blue) Fermi contour. These Fermi contours result from the two helicity bands of the Rashba metal. In panel (a) the two Fermi surfaces coincide, a pro\-per\-ty which is reflected in the mixed (purple) color coding of the Fermi surface. In panel (b) we have indicated the parameter $\delta k_{\rm soc}$ used in the main text, which defines the splitting of the two helicity Fermi contours.}
\label{fig:fig1}
\end{center}
\end{figure}

\noi where $\chi_{||}^{\rm spin}$ indicates the in-plane spin-spin suscepti\-bi\-lity. At $T=0$, we find $\chi_{||}^{\rm spin}=\chi_\perp^{\rm spin}/2=m_e/2\pi\hbar^2$. This implies that $\Lambda$ can be re-written according to:
\begin{align}
\Lambda\big(0<\delta\ll1\big)=\frac{2m_e\upsilon}{\hbar}\frac{e}{4\pi\hbar}\,.
\end{align} 

Thus, we find the ratio $\Lambda/{\cal X}=\delta k_{\rm soc}$, where we introduced the Rashba SOC wavenumber $\delta k_{\rm soc}=2m_e\upsilon/\hbar$, which yields the SOC splitting of the two helical branches, as shown in Fig.~\ref{fig:fig1}. Therefore, in order for the two mechanisms, i.e., in-plane field $M_{||}$ versus out-of-plane magnetization gradient $\partial M_z$, to lead to the same superconducting phase gradient $|\mathbf{q}_0|$, the following relation needs to be satisfied:
\begin{align}
|M_{||}|\delta k_{\rm soc}=|\partial M_z|\sim \frac{|\delta M_z|}{\xi_{\rm grad}}\,.
\end{align}

From the above, we find the following simple result: what controls the relative ratio of the strengths of these two effects is how the Rashba splitting $\delta k_{\rm soc}$ of the two helicity bands compares to the characteristic length scale $\xi_{\rm grad}$ for which the out-of-plane magnetization is modified by $\delta M_z$. Therefore, if we assume the same order of magnetization energies $|M_{||}|\sim|\delta M_z|$, the magnetization gradient contribution to $|\mathbf{q}_0|$ is do\-mi\-nant only when the following inequality is satisfied: 
\begin{align}
\xi_{\rm grad}<1/\delta k_{\rm soc}\,.    
\end{align}

\noi The above reveals that the value of $\xi_{\rm grad}$ required to sa\-ti\-sfy the above scales inversely proportional to the Rashba SOC strength $\upsilon$.

\section{Lattice Model and method for numerical approach}\label{latticebasics}

In this section we describe the lattice model and the method used to determine the supercurrents and the current-phase relations. The Hamiltonian without magnetization effects is given by:
\begin{align}
\mathcal{H} &= \mathcal{H}_{\rm kin} + \mathcal{H}_{\rm soc} + \mathcal{H}_{\rm SC},\label{eq:Hamiltonian}
\end{align}
where
\begin{align}
\mathcal{H}_{\rm kin} &= -t\sum_{\langle \iv,\jv \rangle,\sigma}c_{{\iv}\sigma}^{\dagger}c_{{\jv}\sigma} - \mu\sum_{{\iv},\sigma}c_{{\iv}\sigma}^{\dagger}c_{{\iv}\sigma}, \\ 
\mathcal{H}_{\rm soc} &= -\frac{\lsoc}{2}\sum_{{\iv}}\left[ (c_{{\iv}-\hx\downarrow}^{\dagger}c_{{\iv}\uparrow} - c_{{\iv}+\hx\downarrow}^{\dagger}c_{{\iv}\uparrow}) \right. \nonumber \\ &+ \left. i(c_{{\iv}-\hy\downarrow}^{\dagger}c_{{\iv}\uparrow} - c_{{\iv+\hy}\downarrow}^{\dagger}c_{{\iv}\uparrow}) + \rm{H.c.}\right], \\ 
\mathcal{H}_{\rm SC} &= \sum_{{\iv}}\Delta_{{\iv}}(c_{{\iv}\uparrow}^{\dagger}c_{{\iv}\downarrow}^{\dagger} + \rm{H.c.}),
\label{Equation:Tight_binding_Hamiltonian}
\end{align}

\noi  corresponds to the kinetic term, the Rashba SOC and the superconducting term, respectively. In the above equations, $c_{\mathbf{i}\sigma}^{\dagger} (c_{\mathbf{i}\sigma})$ refer to the electronic creation (annihilation) operator, where $\iv$ is used as the shorthand notation for $\Rv_i$ denoting the coordinates on a 2D square lattice (lattice constant $a=1$) and $\sigma$ refers to the spin. We include  nearest-neighbor (NN) hopping $t=1$ and a chemical potential $\mu$. Further, $\lsoc$ and $\Delta_\iv$ denote the strength of the Rashba SOC and the superconducting order parameter. Below we neglect any contributions from the electromagnetic vector potential in the lattice studies, assuming that such effects are negligible. Possible consequences of their inclusion can be inferred from previous studies~\cite{Nakamura_orbital2023}.

Using the spinor $\Psi_\iv^\dag = (c_{\iv \up}^\dag, c_{\iv\down}^\dag, c_{\iv \up}, c_{\iv\down})$, we construct the $4N^2 \times 4N^2$ Bogoliubov-de Gennes (BdG) Hamiltonian and solve self-consistently at each site for the superconducting order parameter:
\begin{equation}
    \Delta_\iv = V_{\rm SC} \left( \la c_{\iv\up} c_{\iv\down} \ra - \la c_{\iv\down} c_{\iv\up} \ra \right),
\end{equation}
assuming an on-site pairing interaction $V_{\rm SC}=-1.0$ resulting in $\Delta=0.2$ for the homogeneous system in the absence of external fields and with the typical values of $\mu=-1$ and $\lambda_{\rm soc}=0.2$ used below. 

We compute the current densities between all NN bonds, which have two contributions due to the presence of Rashba SOC:
\begin{equation}
     J_{\iv,\hdelta}  =  J_{\iv,\hdelta}^t + J_{\iv,\hdelta}^{\rm soc},
    \label{eq:bondJ}
\end{equation}
where $\hdelta = \{\hx,-\hx,\hy,-\hy\}$ denote the four NN bonds to site $\iv$. The first contribution to the current density is derived from the hopping term of the Hamiltonian and can be written as:
\begin{align}
    J_{\iv,\hdelta}^t &= it \sum_\sigma \la c^\dagger_{\iv + \hdelta \sigma} c_{\iv \sigma}  -  \rm{H.c.}\ra.
    \label{eq:currentt}
\end{align}
The latter term in Eq.~\eqref{eq:bondJ} due to SOC is given by:
\begin{align}
     J_{\iv,\pm \hx}^{\rm soc}  &= i \frac{\lsoc}{2} \sum_{\iv,\sigma} \la \pm \sigma c^\dagger_{\iv \pm \hx \sigma} c_{\iv \Bar{\sigma}} - \rm{H.c.} \ra, 
     \label{eq:currentlx} \\
     J_{\iv,\pm \hy}^{\rm soc} &= \pm \frac{\lsoc}{2} \sum_{\iv,\sigma}  \la c^\dagger_{\iv \pm \hy \sigma} c_{\iv \Bar{\sigma}}  -  \rm{H.c.} \ra,
     \label{eq:currently}
\end{align}
with $\Bar{\sigma} = - \sigma$.
Finally, the total current operator on each site is defined as the average vector of the two adjacent bonds:
\begin{align}
     {\bf{J}}_\iv  = \frac{1}{2} \sum_{\hdelta} \hdelta J_{\iv,\hdelta}.
\end{align}
The current expressions Eqs.~(\ref{eq:currentt})-(\ref{eq:currently}) contain an additional factor of $e/(\hbar a^2)$, which enters our current unit in the following.

The Hamiltonian for the in-plane magnetic field in the $y$ direction is given by:
\begin{equation}
    \mathcal{H}_{B} = - B_y \sum_{\iv,\sigma,\sigma'} (\sigma_y)_{\sigma\sigma'} c_{{\iv}\sigma}^{\dagger}c_{{\iv}\sigma'}.
\end{equation}
In contrast, the out-of-plane magnetization gradient case is modeled by:
\begin{equation}\label{MagnH}
    \mathcal{H}_{M_z} = -\sum_{{\iv},\sigma,\sigma'} M_{z,{\iv}} (\sigma_z)_{\sigma\sigma'} c_{{\iv}\sigma}^{\dagger}c_{{\iv}\sigma'},
\end{equation}
with $M_{z,\iv}$ a site-dependent magnetization.

\begin{figure}[t!]
    \begin{center}
   	\includegraphics[angle=0,width=7cm]{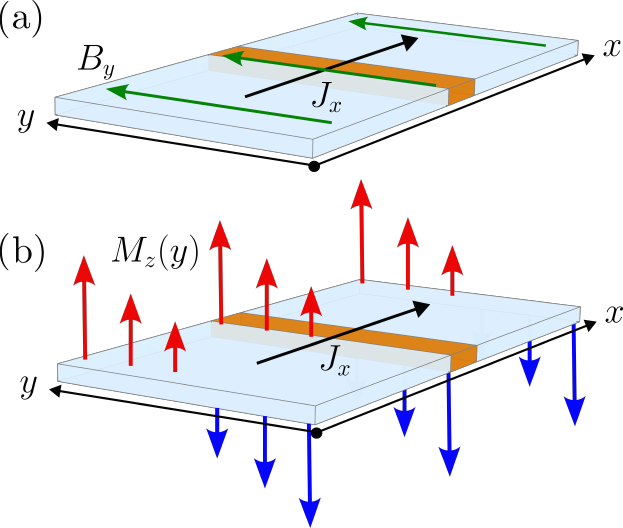}
    \caption{Illustration of the Josephson junction devices for the cases of (a) an in-plane magnetic field and (b) an out-of-plane magnetization gradient. The orange region corresponds to the weak link junction area, and $J_x$ denotes the current arising in the $x$ direction.}
    \label{fig:fig2}
    \end{center}
\end{figure}

In order to compare the helical phase and the superconducting phase gradient for both field configurations, we first impose open boundary conditions (OBC) and solve self-consistently for the order parameter at each site. To obtain the current-phase relations, we impose a phase dif\-fe\-ren\-ce $\varphi$ between both ends of the system at each iteration of the self-consistency. Insisting on a phase gradient $\varphi$ across the junction in each iteration induces a current, which, only when full self-consistency is obtained at each lattice site, respects charge conservation~\cite{Andersen05,Andersen06}.

\section{Results from the lattice model}\label{sec:latticeresults}

The setup for the Josephson junction is shown in Fig.~\ref{fig:fig2}. As seen it consists of a symmetric junction with a weak link of reduced transparency (orange middle ribbon) modelled by two columns of lattice bonds of hopping $t'$. In this section, we first discuss helical superconductivity, i.e., the case of no bulk currents and additionally $t'=t$ (no weak link). Afterwards, we turn to the current-carrying states with imposed phase difference $\varphi$ across the junction, and discuss both the superconducting diode effect ($t'=t$) and the Josephson diode effect ($t'<t$).

\begin{figure}[t!]
\begin{center}    \includegraphics[angle=0,width=0.95\linewidth]{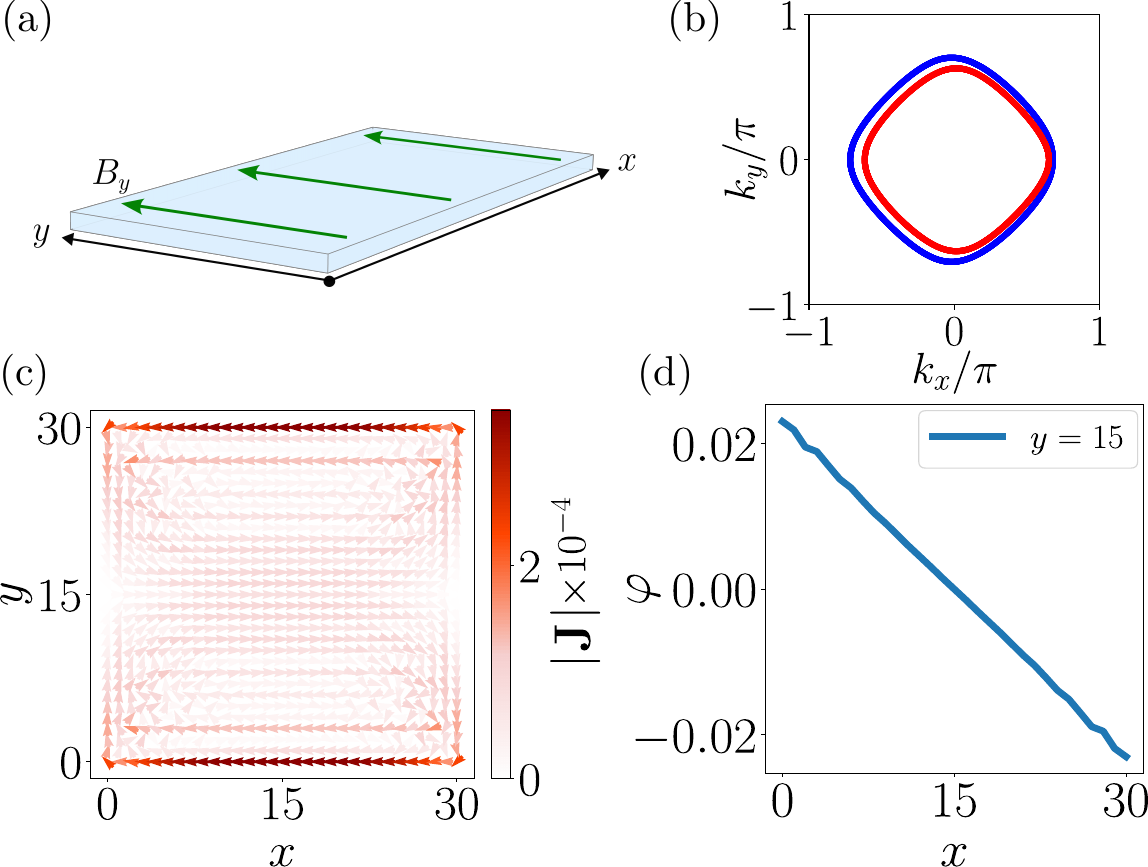}
    \caption{(a) Helical superconductivity generated in a Rashba superconducting strip in a transverse in-plane magnetic field. (b) Fermi surface for $\lsoc = 0.2$, $B_y = 0.1$ and $\mu = -1.0$. (c) Example of the remnant total currents at each bond on a $31 \times 31$ system after requiring selfconsistency. All currents are negligible and vanish in the thermodynamic limit except near some edges. (d) Superconducting phase at $y=15$ as a function of position along the $x$-axis, displaying the finite constant gradient of the helical ground state.}
    \label{fig:fig3}
    \end{center}
\end{figure}

\subsection{Helical superconductivity}

\subsubsection{Case of in-plane magnetic field}

Here, we review the helical ground state properties in the case of a superconducting Rashba metal in an in-plane magnetic field, as illustrated in Fig.~\ref{fig:fig3}(a). In the presence of SOC, an in-plane magnetic field shifts the center of the two Fermi surfaces, as shown in Fig.~\ref{fig:fig3}(b). The superconducting ground state adiabatically enters the helical phase where the order parameter, as mentioned in the introduction, acquires a spatial variation of the form $\Delta (\Rv) = \Delta_0 e^{i  \mathbf{q}_0 \cdot \Rv}$~\cite{Samokhin04,Agterberg2007,Kaur2005,Yuan22}.
The ground state carries zero total current due to the perfect compensation of $ J_{\iv,\hdelta}^t $ and $J_{\iv,\hdelta}^{\rm soc} $. We have verified these properties by numerical selfconsistent calculations of the order parameter and the currents. Figure~\ref{fig:fig3}(c) and \ref{fig:fig3}(d) explicitly show the absence of bulk currents and a finite ground state phase gradient, respectively. We have checked that for large systems (or imposing periodic boundary conditions), the bulk currents indeed vanish identically. For finite sized systems as in Fig.~\ref{fig:fig3}(c), the currents may develop inte\-re\-sting patterns with non-vanishing contributions prevalent near some of the sample edges, in this case the edges perpendicular to the magnetic in-plane field direction. The numerically obtained value for the Cooper pair momentum of the helical state presented in Fig.~\ref{fig:fig3} can be shown to be quantitatively consistent with a corresponding free-energy analysis.

\begin{figure}[t!]
    \begin{center}
    \includegraphics[angle=0,width=.95\linewidth]{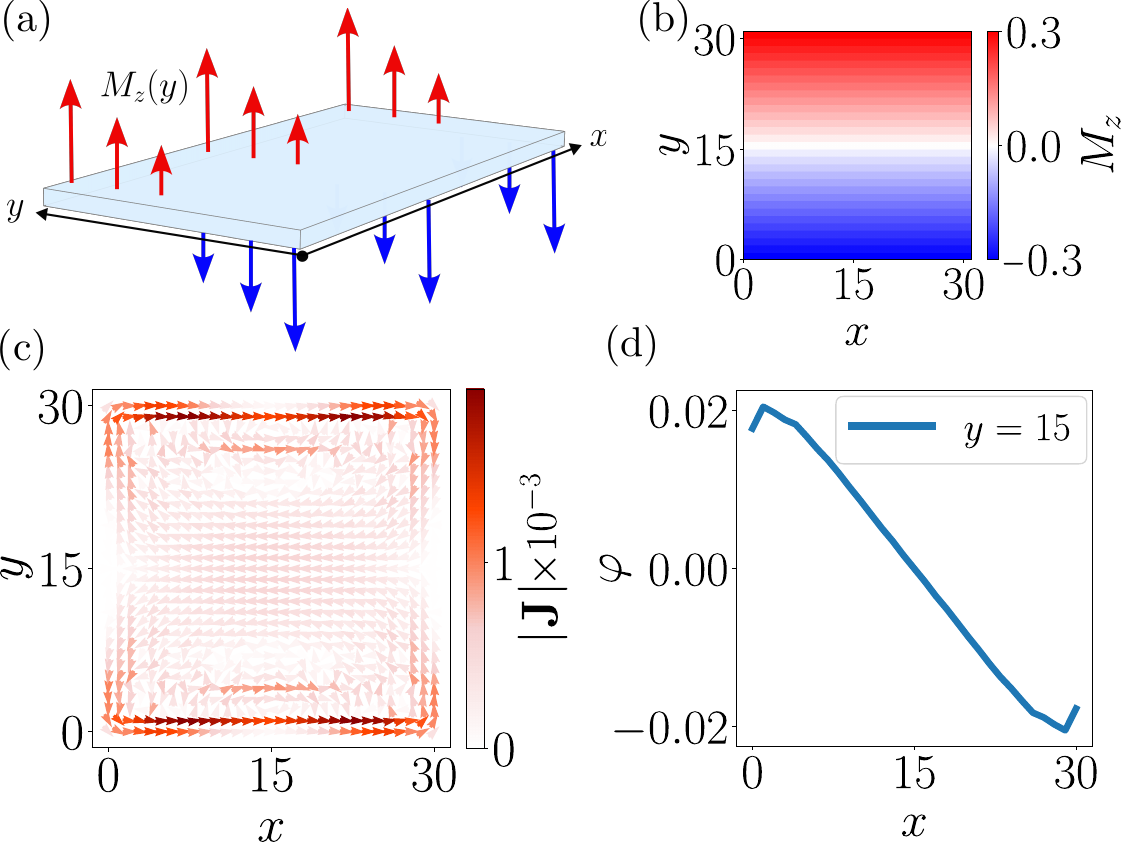}
    \caption{(a) Helical superconductivity generated in a Rashba superconducting strip with an out-of-plane magnetization gradient, as shown in panel (b). (c) Example of the remnant total currents at each bond on a $31 \times 31$ system after re\-qui\-ring self-consistency, with $\lsoc = 0.2$, $2 M_z^{\textrm{max}}/\xi_{\textrm{grad}} = 0.02$ and $\mu = -1.0$. All currents are negligible and vanish in the thermodynamic limit except near the edges. (d) Superconducting phase at $y=15$ as a function of position along the $x$-axis, exhibiting the constant gradient of the helical ground state.}
    \label{fig:fig4}
    \end{center}
\end{figure}

\subsubsection{Case of out-of-plane magnetization gradient}

For helical superconductivity originating from a spatially varying out-of-plane magnetization, see Eqs.~(\ref{eq:Jmag}) and (\ref{eq:phasefromJmag}), we focus initially on the simplest case with a uniform magnetization gradient as shown schematically in Fig.~\ref{fig:fig4}(a) and quantitatively in Fig.~\ref{fig:fig4}(b). Figure~\ref{fig:fig4}(c) and \ref{fig:fig4}(d) display the current pattern in the ground state of a $31 \times 31$ system and the associated ground state superconducting phase gradient, respectively. The latter confirms that indeed a helical state is favored also in this case. The total currents in the ground state are again vanishing except for edge effects. 

In Sec.~\ref{sec:caseofrashba} we used the continuum model in the quasiclassical limit to derive the result that in-plane fields lead to a current $J={\rm C}|M_{||}|\delta k_{\rm soc}$ and that out-of-plane magnetization gradients lead to a current $J={\rm C}|\delta M_z|/\xi_{\rm grad}$, with the same proportionality constant ${\rm C}$. Hence, for helical superconductivity, $B_y \delta k_{\rm soc}$ and $2 M^{\rm max}_z/\xi_{\rm grad}$ should generate comparable phase gradients. This result can be explicitly verified within the lattice approach. For this purpose, we selfconsistently solve for the supercon\-duc\-ting gap imposing OBC for an in-plane field and an out-of-plane magnetization gradient. Since the ground state enters the helical phase in both cases as shown above, the superconducting order pa\-ra\-me\-ter develops a phase gradient $\partial \varphi/ \partial x$ in the $x$ direction, which we evaluate numerically. 

\begin{figure}[t!]
    \begin{center}
\includegraphics[angle=0,width=.85\linewidth]{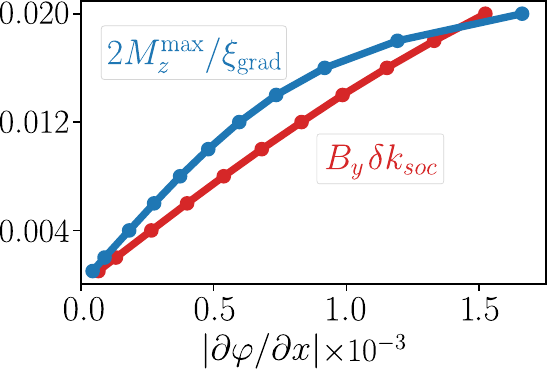}
    \caption{Comparison of the superconducting phase gradient generated in the Rashba superconductor by the in-plane field $B_y \, \delta k_{soc}$ (red curve) or an out-of-plane magnetization gradient $2M_z^{\rm max}/\xi_{\rm grad}$ (blue curve), in the case of $\lsoc = 0.2$, $\mu = -1.0$ and system size $31\cross 31$. An equi\-va\-lent helical superconducting phase is generated by the two dif\-fe\-rent means.}
    \label{fig:fig5}
    \end{center}
\end{figure}

In Fig.~\ref{fig:fig5} we compare the phase gradients in the helical phases. As seen, the superconducting phase gradients indeed agree semi-quantitatively, par\-ti\-cu\-lar\-ly in the li\-near regime where $|\partial \varphi/ \partial x|$ is small. We attribute the de\-via\-tion to the fact that the finite-size lattice calculations are not sufficiently deep in the quasiclassical regime since $\Delta/E_F\simeq 0.07$ and $E_{\rm soc}/E_F\simeq 0.1$.  As the out-of-plane magnetization becomes larger, the superconducting order becomes increa\-singly inhomogeneous, leading to the non-linear regime where the phase gradient generated by the out-of-plane magnetization crosses the one from the in-plane field. Another possible origin of the arising discrepancy can be the non-circular character of the Fermi lines obtained in the investigated regime.

\subsection{Superconducting and Josephson diode effects}

\begin{figure}[b]
    \begin{center}
    \includegraphics[angle=0,width=.8\linewidth]{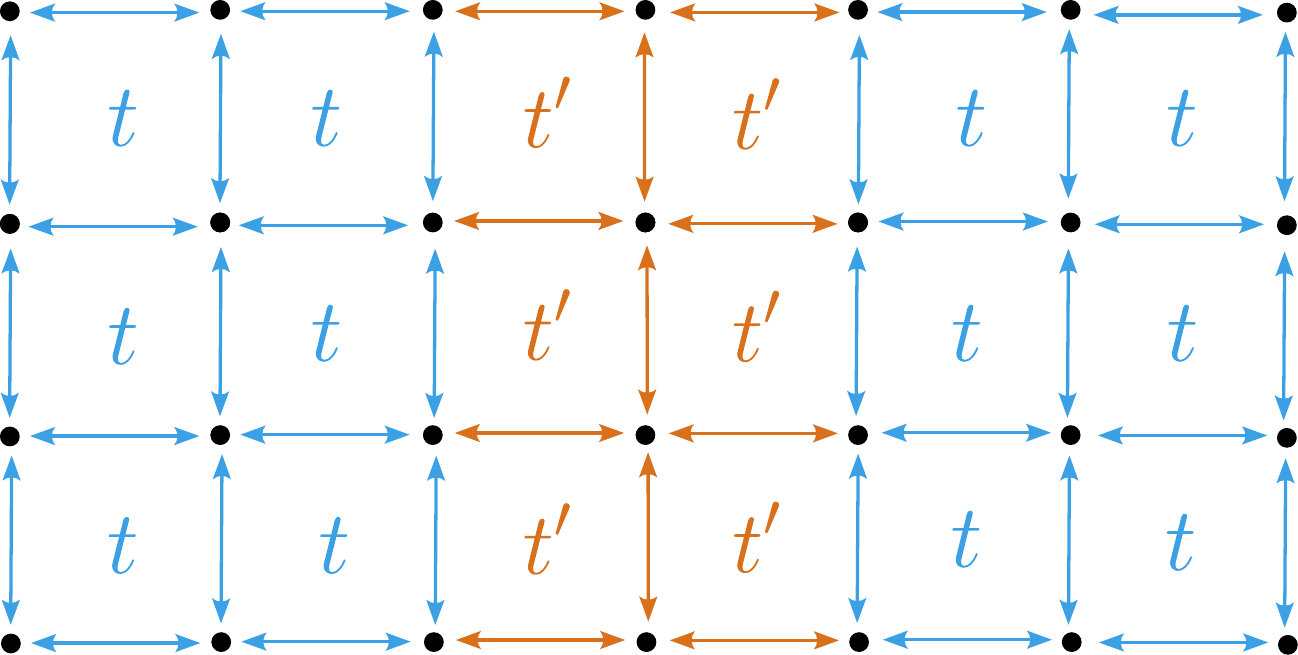}
    \caption{Illustration of the junction within the lattice model; two connected columns of bonds experience hopping of $t'$ as opposed to all other bonds exhibiting hopping integral of $t$.}
    \label{fig:fig6}
    \end{center}
\end{figure}

In this section we turn to a discussion of the current-carrying states in Josephson junctions. The currents are introduced by an enforced phase difference across the junction, which can be implemented numerically by insisting on such a phase difference in the iteration process~\cite{Andersen05}. In the absence of magnetization gradients or external fields, the current-phase relations exhibit the expected form (not shown), i.e., saw-tooth shaped curves that get deformed into sinusoidal current-phase relations as $t'$, and thereby the junction transparency, is reduced. Figure~\ref{fig:fig6} shows explicitly the model for the junction and which bonds experience the modified hopping integral $t'$.

\begin{figure}[tb]
    \begin{center}
    \includegraphics[angle=0,width=.95\linewidth]{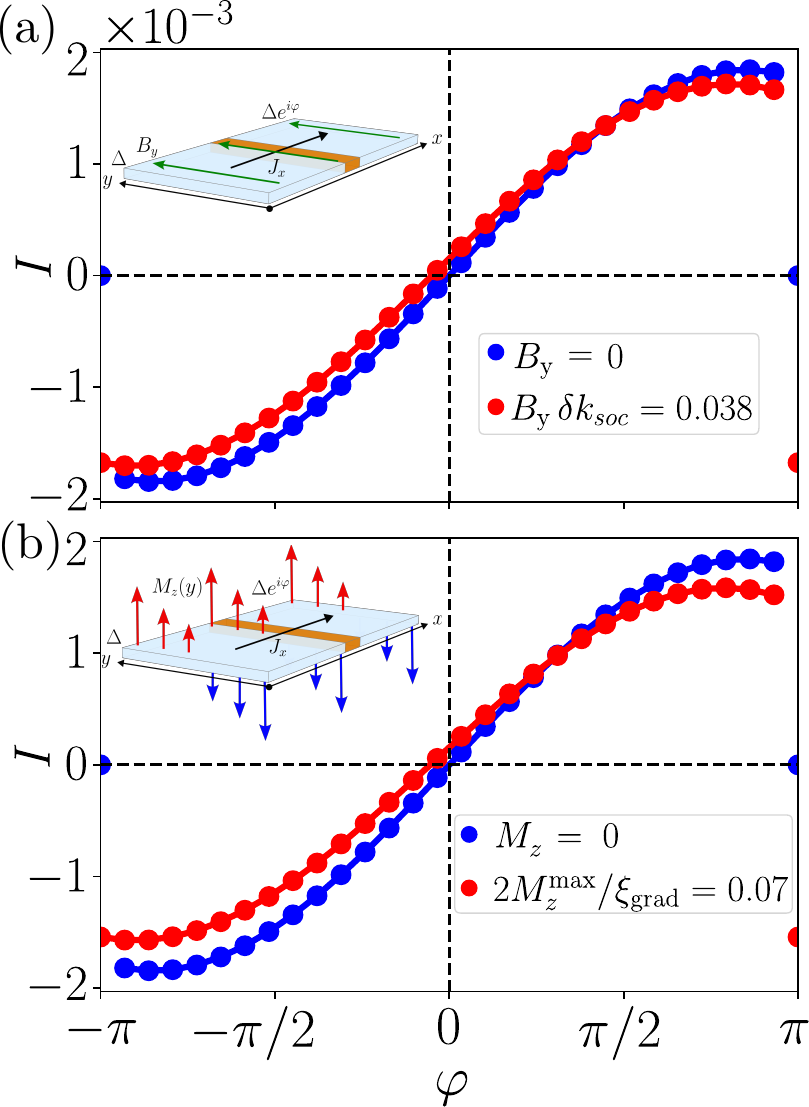}
    \caption{Current-phase relations for the case of (a) in-plane magnetic field and (b), out-of-plane magnetization gradient with $\lsoc = 0.2$, $\mu = -1.0$, $t' = 0.1$ and a system size of $21\cross 11$. The red (blue) curves indicate the cases with (without) magnetic field/magnetization gradient where the current-phase relations are nonsymmetric (symmetric).}
    \label{fig:fig7}
    \end{center}
\end{figure}

In the presence of the in-plane field or out-of-plane magnetization gradient, the system enters the helical state. The inherent directionality dictating the helical superconducting phase can be uncovered in the current-carrying state. This is evident from shifted and asymmetric current-phase relations and the associated diode effect of the junctions. For the case $t'=t$ we find that both setups produce a weak superconducting diode effect. In the following we focus mainly of the Josephson diode effect with reduced junction transparency. Figure~\ref{fig:fig7} displays examples of typical current-phase relations in the helical state with $t'=0.1$ both in the case of an in-plane magnetic field Fig.~\ref{fig:fig7}(a) and an out-of-plane magnetization gradient Fig.~\ref{fig:fig7}(b). As seen, the Josephson current indeed exhibits a diode behavior. Below we elaborate on the diode efficiency of the Josephson junctions.

\begin{figure}[t!]
    \begin{center}
    \includegraphics[angle=0,width=.85\linewidth]{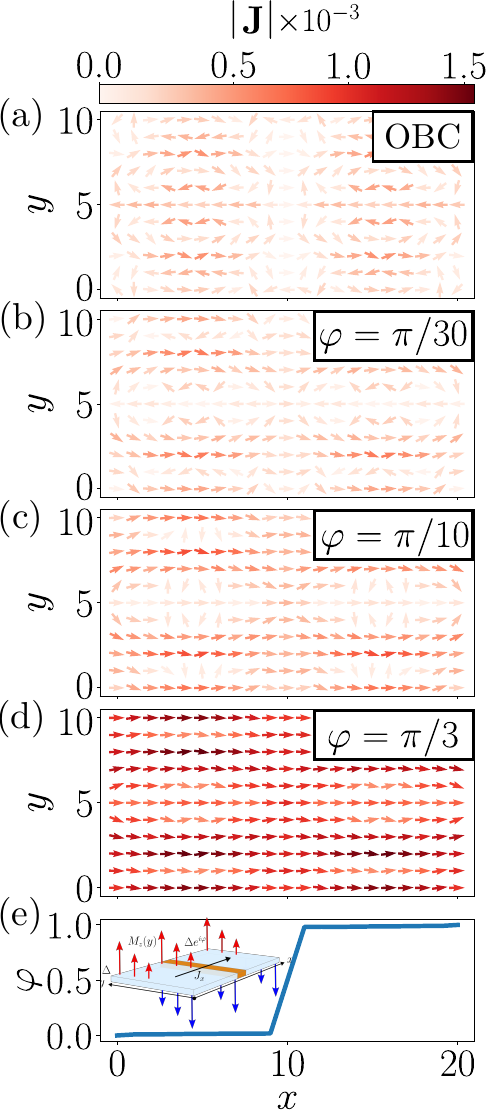}
    \caption{Evolution of current patterns in the case of a homogeneous out-of-plane magnetization gradient $2 M_z^{\rm max}/\xi_{\rm grad} = 0.04$ for (a) OBC, and imposed phase gradients of (b) $\varphi=\pi/30$, (c) $\varphi=\pi/10$, and (d) $\varphi=\pi/3$, in  the case of $\lsoc = 0.2$, $\mu = -1.0$, $t' = 0.1$ and a system size of $21\cross 11$. For the latter case (d), the resulting superconducting phase is shown in panel (e) for $y=5$, displaying the phase drop in the junction (orange region in the inset). For the case in panel (a) all bulk currents vanish as the system size increases, similar to Fig.~\ref{fig:fig4}(c).}
    \label{fig:fig8}
    \end{center}
\end{figure}

In Fig.~\ref{fig:fig8} we show the spatially resolved current profile across the junction in the presence of an out-of-plane magnetization for the same case as in Fig.~\ref{fig:fig7}(b), i.e., a representative Josephson diode case reasonably close to the tunneling limit with $t'=0.1$. For the case of OBC shown in Fig.~\ref{fig:fig8}(a) there are remnant current loops visible due to the finite size of the system. They, however, cancel identically across any cross-section of the sample, i.e., there is no net current flow in any direction. These edge currents are simply a consequence of the relatively small system sizes under consideration. As evident from Fig.~\ref{fig:fig7}(b) and explicitly shown in Fig.~\ref{fig:fig8}, however, a fixed phase gradient leads to a net total current flow across the junction. The current direction can be switched by the sign of the magnetization gradient or by the imposed phase difference.

\begin{figure}[tb]
    \begin{center}  
    \includegraphics[angle=0,width=.95\linewidth]{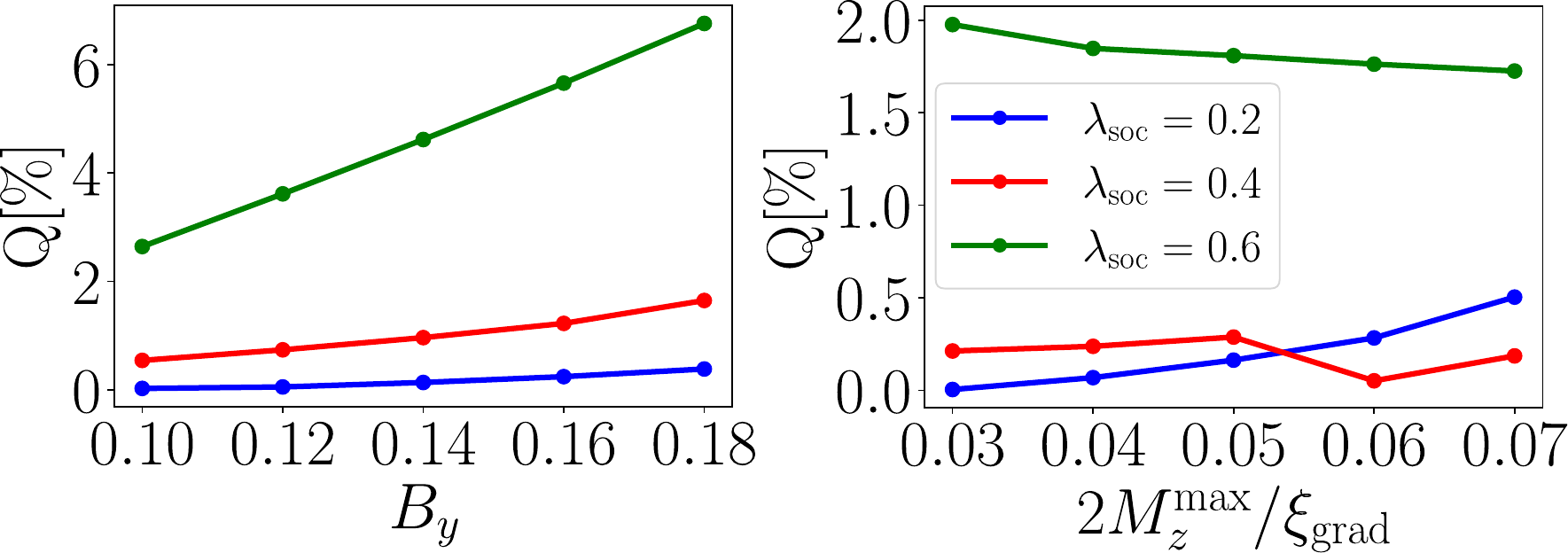}
    \caption{Josephson junction diode efficiency $Q$ for the case of an in-plane magnetic field (a) or an out-of-plane magnetization gradient (b) for different values of the Rashba SOC strength $\lambda_{\rm soc}$, with $\mu = -1.0$, $t'=0.1$ and a system size $21\times 11$. } 
    \label{fig:fig9}
    \end{center}
\end{figure}

Next we turn to a discussion of the Josephson junction diode efficiency $Q$~\cite{He_2022,Yuan22,Daido22,Bergeret22,Davydova22} defined by:
\begin{equation}
    Q = \frac{\abs{I_{\rm{max}}} - \abs{I_{\rm{min}}}}{\abs{I_{\rm{max}}} + \abs{I_{\rm{min}}}},
\end{equation}
\noi where $I_{\rm{max}}$ and $I_{\rm{min}}$ denote the maximum and minimum currents, respectively. Here, $I$ is the sum of all the currents through one cross-sectional cut normalized by the length in the $y$ direction. We obtain these values from the numerically computed current-phase relations. Ty\-pi\-cal\-ly, the associated diode effects arising from a helical superconducting phase are small, thus implying the same for the corresponding values of $Q$. Nevertheless, the precise outcome  will obviously depend on input parameters which, in the current case, consist of the amplitude of the Rashba SOC, the Fermi surface filling set by the che\-mi\-cal potential, the amplitude of the respective external field and, finally, the transparency of the junction. 

For fixed junction transparency $t' =0.1$, we show respectively in Fig.~\ref{fig:fig9}(a) and \ref{fig:fig9}(b) the dependence of the diode efficiency on the in-plane magnetic field and the out-of-plane magnetization gradient for different values of the SOC parameter $\lambda_{\rm soc}$. As seen, the diode efficiency of the junction may be significantly enhanced by the SOC and the amplitude of the driving magnetic field or magnetization gradient. Too large fields, however, destroy superconductivity and puts a natural bound on $Q$.  

\begin{figure}[t!]
    \begin{center}  
    \includegraphics[angle=0,width=0.99\linewidth]{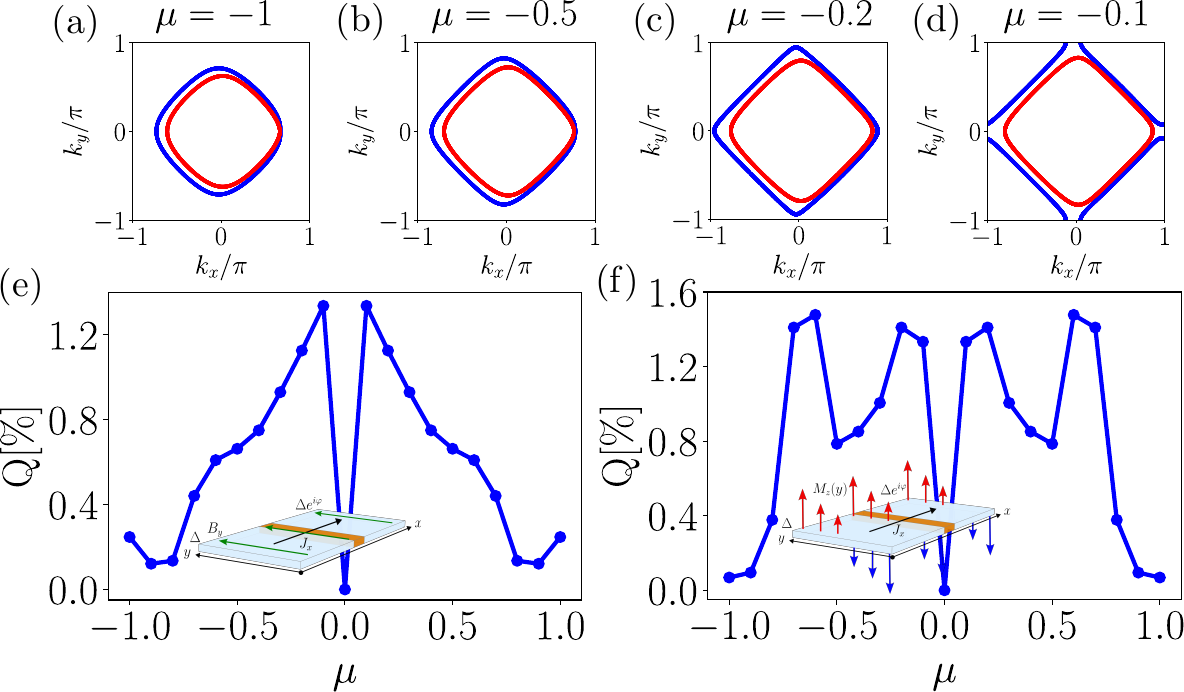}
    \caption{(a)-(d) Fermi surfaces for an in-plane magnetic field,  considering different values of the chemical potential $\mu$, with $\lsoc = 0.2$ and $B_y = 0.16$.  Josephson junction diode efficiency $Q$ as a function of chemical potential $\mu$ in the case of (e) an in-plane field $B_y = 0.16$ and (f) and out-of-plane magnetization gradient $2 M_z^{\rm max}/\xi_{\rm grad} = 0.04$, with $\lsoc = 0.2$, $t'  = 0.1$ and a system size $21\times 11$. 
    } 
    \label{fig:fig20}
    \end{center}
\end{figure}

Aside from the strength of the SOC, we find that the value of the che\-mi\-cal potential $\mu$ also significantly affects $Q$. This can be seen from Fig.~\ref{fig:fig20}, which maps out the chemical potential dependence of $Q$. Notably, our numerical approach allows us to go beyond the quasiclassical regime discussed in   Sec.~\ref{sec:caseofrashba}. Panels \ref{fig:fig20}(a)-(d) display the evolution of the Fermi surface versus $\mu$ for the homogeneous case of a constant in-plane magnetic field. In that case, see Fig.~\ref{fig:fig20}(e), the maximum efficiency arises from one of the split Fermi surfaces acquiring significant contributions from the van Hove singularities. In the inhomogeneous case of an out-of-plane magnetization gradient, the efficiency $Q$ versus $\mu$ is shown in Fig.~\ref{fig:fig20}(f). In that case too, we attribute the $\mu$-dependent efficiency ma\-xi\-ma to regions of favorable density of states conditions. At this point, it is interesting to observe that in the lattice model $Q=0$ for $\mu=0$, which is a property not found for the continuum model defined in Sec.~\ref{sec:caseofrashba}, as shown in Ref.~\onlinecite{He_2022}. This qualitative difference is due to a symmetry that emerges for the lattice model when $\mu=0$. This symmetry is easier to describe using the BdG matrix Hamiltonian $\hat{{\cal H}}(\mathbf{k})$ obtained after expressing Eq.~\eqref{eq:Hamiltonian} in reciprocal space. In this space, the arising symmetry is effected by the following shift operation of the wave vector $\mathbf{k}\mapsto\mathbf{k}+(\pi,\pi)$ accompanied by an exchange between time-reversed electron and hole partners.

From the above results, it is evident that we find overall comparable efficiencies for the two diode mechanisms, that is, for in-plane field versus out-of-plane magnetization gradient. However, the in-plane field has the disadvantage that large amplitudes of it eventually destroys superconductivity through pair-breaking. In stark contrast, while this effect may also happen for the gradient case, this would still require large magnetization amplitudes. This is because for the magnetization gradient profiles considered here, the net magnetic moment is zero, thus rendering such setups more compatible with the superconducting elements of the circuit. Hence, this poses yet another interesting possibility; to engineer the spatial structure of the out-of-plane magnetization in order to optimize the diode efficiency.

\begin{figure}[t!]
    \begin{center}
    \includegraphics[angle=0,width=.95\linewidth]{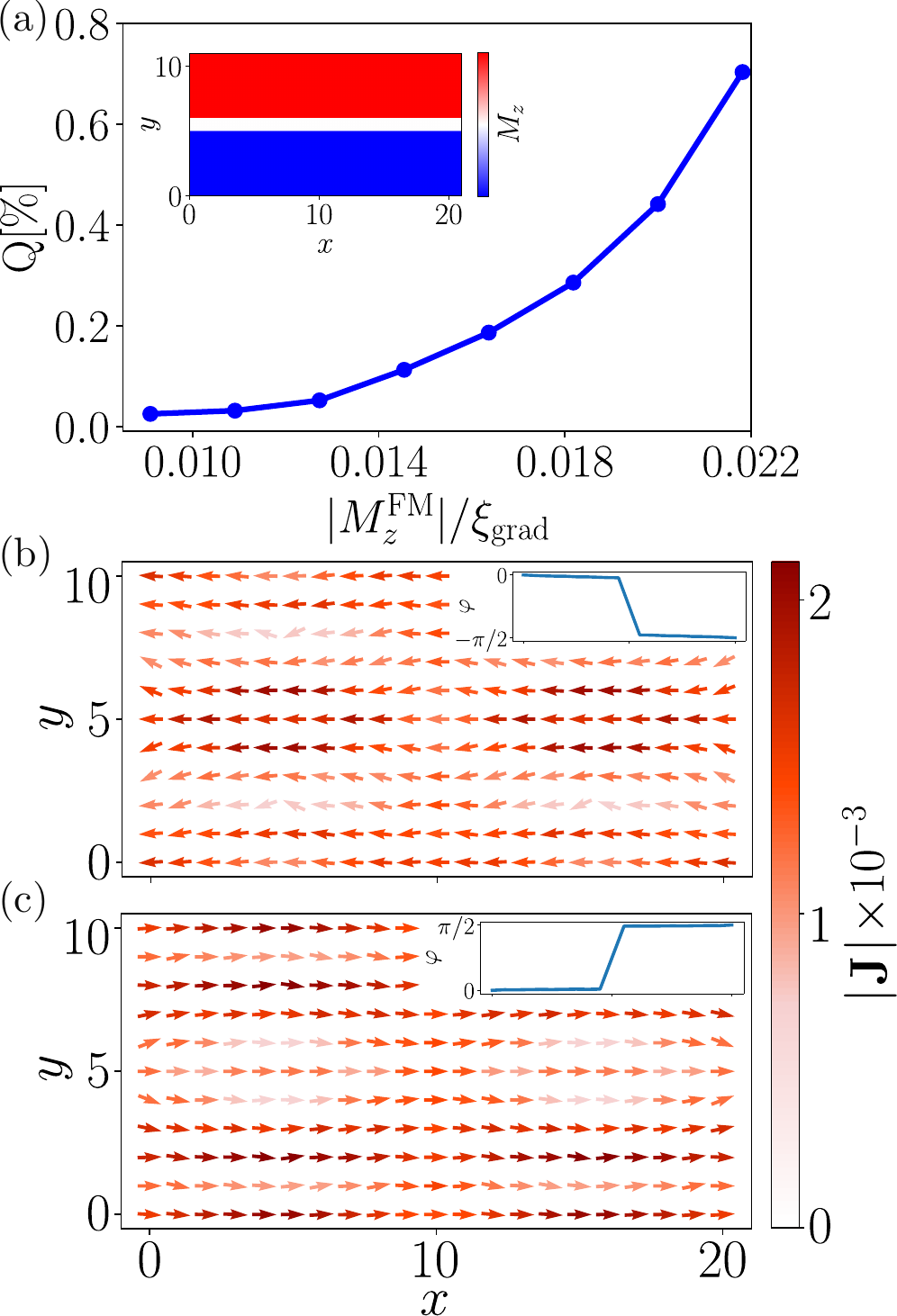}
    \caption{(a) Josephson diode efficiency $Q$ versus the amplitude of the magnetization of the ferromagnetic regions, in the case of a $21\cross 11$ system, $\lsoc = 0.2$, $\mu = -1.0$, $\xi_{\rm grad}=11$ and $t'  = 0.1$. In this case the Rashba superconductor is exposed to a ferromagnetic domain wall with the magnetization as shown in the inset. Panels (b) and (c) show the current distribution for opposite phases biases across the sample. One observes that the current flow is not restricted to the domain wall region. Remarkably, however, the polarity of the phase bias can toggle between high and low domain wall current.
    }
    \label{fig:fig21}
    \end{center}
\end{figure}

Here, however, we do not pursue this optimization problem, but instead aim at unveiling an alternative related setup, which appears equally prominent and feasible to rea\-li\-ze in experiments. This is a ferromagnetic (FM) domain wall junction. A sketch of such a platform is shown in the inset of Fig.~\ref{fig:fig21}(a). In a similar fa\-shion to the previously examined magnetic configurations, the FM junction also generates a helical ground state and supports therefore a Josephson diode effect. 

From the current-phase relations, we have again extracted the directional-dependent critical currents and the associated efficiencies shown in Fig.~\ref{fig:fig21} as a function of the amplitude of the out-of-plane magnetization. For the magnetization used in Fig.~\ref{fig:fig21} superconducti\-vi\-ty still persists throughout the junction. As seen from Fig.~\ref{fig:fig21}(a), $Q$ increases with the amplitude of the out-of-plane FM, until eventually this field becomes too large and it suppresses superconductivity. 

From Figs.~\ref{fig:fig21}(b) and~\ref{fig:fig21}(c), we observe that the resulting charge currents are not restricted to flow in close vicinity of the domain wall. Instead they extend over the entire transverse dimension of the sample. The above property of the currents precisely reflects that transport stems from the charged condensate extending throughout the entire junction. Indeed, from studying the energy spectra, we also confirm that for the pa\-ra\-me\-ter window stu\-died here there exist no domain wall boundary quasiparticle modes that could contribute to transport. We remark, however, that this fin\-ding could not have been taken for granted, since such a possibility generally exists for the model of Eq.~\eqref{eq:Hamiltonian} in the presence of a ferromagnetic domain wall. Akin to previous theoretical predictions for Rashba superconductors~\cite{Sau2010,QiHughesZhang}, also here a topological superconductor in class D can arise on each side of the wall when $|M_z|$ takes values in a desired range. For the NN hoppings and Rashba SOC considered here, along with the assumptions $\Delta>0$ and $|\mu|<t=1$, each side of the domain wall with net magnetization $\pm|M_z|$ behaves as a topological superconductor with a Chern number ${\cal N}=\pm2$ for $\Delta<|M_z|<\sqrt{\Delta^2+(4t-|\mu|)^2}$ and ${\cal N}=\pm1$ for $\sqrt{\Delta^2+(4t-|\mu|)^2}<|M_z|<\sqrt{\Delta^2+(4t+|\mu|)^2}$, respectively. The emergence of such topological states gives rise to correspondingly $4$ and $2$ chiral Majorana edge modes which are trapped at the domain wall. The properties of the diode effect arising for such topological scenarios are interesting on their own and therefore deserve a de\-di\-ca\-ted study. However, we leave this for a future study, since investigating these boundary mode contributions go beyond the scope of the current work. 

We conclude this section with possible optimization strategies of the diode efficiency for the present setup. Similar to the cases studied earlier, $Q$ can be also here further enhanced by appropriate tuning of the values for the parameters $\lambda_{\rm soc}$ and $\mu$. For example, using $\lambda_{\rm soc}=0.6$ and $\mu=-0.2$ one obtains $Q\simeq 3\%$ for the single FM domain wall junction. Another optimization approach which promises an enhanced diode effect versatility is the employment of layered ferromagnet-semiconductor-superconductor heterostructures, e.g., similar to the Majorana multi-layer platforms discussed in Ref.~\onlinecite{Sau2010}. In such hybrid systems, the Rashba SOC appears in the semiconductor element, which additionally experiences an exchange field and a pairing gap due to proximity effects. Since the pairing gap in the effective Rashba superconductor which becomes engineered in the semiconductor is no longer driven by interactions, but is instead externally imposed, it appears feasible that such a setup  allow for higher efficiencies.

\section{Magnetization-gradient diodes in the absence of spin-orbit coupling}\label{sec:NoSOC}
Up to this point, we have explored superconducting platforms which feature Rashba SOC. It is the pre\-senc\-e of the latter that has allowed us to solely consider a magnetization gradient pointing only along the out-of-plane axis. In the case that the involved superconducting material is free of any SOC which violates inversion symmetry, a diode effect from a spatially inhomogeneous magnetization field remains possible, albeit it requires a more complex magnetization profile.

Specifically, in the case of 2D systems discussed here, the magnetization field needs to consist of an additional contribution which generates effectively the required Rashba SOC that we considered in the pre\-ce\-ding analysis. This part may have the form of a magnetic texture crystal~\cite{MostovoyFerro,Christensen_18,KotetesSPIE}, i.e., a periodically re\-pea\-ting magnetic texture in space, which is known to give rise to a synthetic Rashba SOC~\cite{BrauneckerSOC,KarstenNoSOC,Ivar,KlinovajaGraphene,Nakosai2013,Kontos}. Whether the spatial profile $\bm{m}(\bm{r})$ of a magnetization texture crystal effectively leads to a Rashba SOC of the general form $\hat{\bm{n}}\cdot\bm{p}\bm{W}_{\hat{\bm{n}}}\cdot\bm{\sigma}$, can be judged from the structure of the vectorial coefficients $\bm{W}_{\hat{\bm{n}}}$ which are given by~\cite{Christensen_18,KotetesSPIE}:
\begin{align}
\bm{W}_{\hat{\bm{n}}}\propto \int_{\rm MUC}dV\,\bm{m}(\bm{r})\times\frac{\partial\bm{m}(\bm{r})}{\partial\bm{n}},    
\end{align}

\noi where $\partial/\partial\bm{n}\equiv\hat{\bm{n}}\cdot\partial/\partial\bm{r}$ denotes the directional derivative in coordinate space in the direction defined by the unit vector $\hat{\bm{n}}$. The integral employed above is over the volume of the magnetic unit cell (MUC), which is determined by the periodicity dictating the magnetic texture.

In certain cases, it is straightforward to identify the quantities $\bm{W}_{\hat{\bm{n}}}$ from the microscopic model. As an example, consider for instance the continuum model discussed in Sec.~\ref{sec:caseofrashba}, with the difference that there is no Rashba SOC, but the system is instead under the influence of a magnetic helix crystal $\bm{m}(x)=m\big(\sin({\cal Q}x),0,\cos({\cal Q}x)\big)$. Due to the fact that $|\bm{m}(\bm{r})|$ is spatially uniform, one can carry out a spin-dependent gauge transformation~\cite{BrauneckerSOC}, and find an effective anisotropic Rashba SOC of the form $\upsilon_{\rm eff}p_x\sigma_y$ with $\upsilon_{\rm eff}=\hbar {\cal Q}/2m_e$. Thus, in this case, one can readily set $\bm{W}_{\hat{\bm{n}}}=\int_{\rm MUC}dV\,\hat{\bm{m}}(\bm{r})\times\partial\hat{\bm{m}}(\bm{r})/\partial\bm{n}$ with $\hat{\bm{m}}(\bm{r})=\bm{m}(\bm{r})/|\bm{m}(\bm{r})|$. Hence, for the magnetic helix profile, one finds that the only nonzero vectorial coefficient is given by $\bm{W}_{\hat{\bm{x}}}=(0,{\cal Q},0)$. Consider now instead a helix of the form $\bm{m}(y)=m\big(0,\sin({\cal Q}y),\cos({\cal Q}y)\big)$. This, in turn, generates the Rashba SOC term $-\upsilon_{\rm eff}p_y\sigma_x$.

It now becomes clear that a possible minimal magnetic texture crystal configuration that can generate a synthetic Rashba SOC of the form $\upsilon_{\rm eff}(p_x\sigma_y-p_y\sigma_x)$ is of the so-called spin-whirl crystal (SWC) form~\cite{Christensen_18}:
\begin{align}
\bm{m}_{\rm swc}(\bm{r})=m\big(f\sin({\cal Q}x),f\sin({\cal Q}y),\cos({\cal Q}x)+\cos({\cal Q}y)\big)\,,\label{eq:SWC}    
\end{align}

\noi with an anisotropy factor $f$ which is generally not equal to unity. While the link of the vectorial coefficients $\bm{W}_{\hat{\bm{n}}}$ is no longer as simple as in the case of the single magnetic helix crystal profile, one still expects the strength of the effective Rashba SOC to appro\-xi\-ma\-te\-ly continue to be given by $\upsilon_{\rm eff}=\hbar {\cal Q}/2m_e$. As a result, we expect the earlier defined parameter $\delta k_{\rm soc}$ dictating the splitting of the helicity bands to be roughly equal to the magnetic wave number ${\cal Q}$.

\begin{figure}[t!]
    \begin{center}
    \includegraphics[angle=0,width=.9\linewidth]{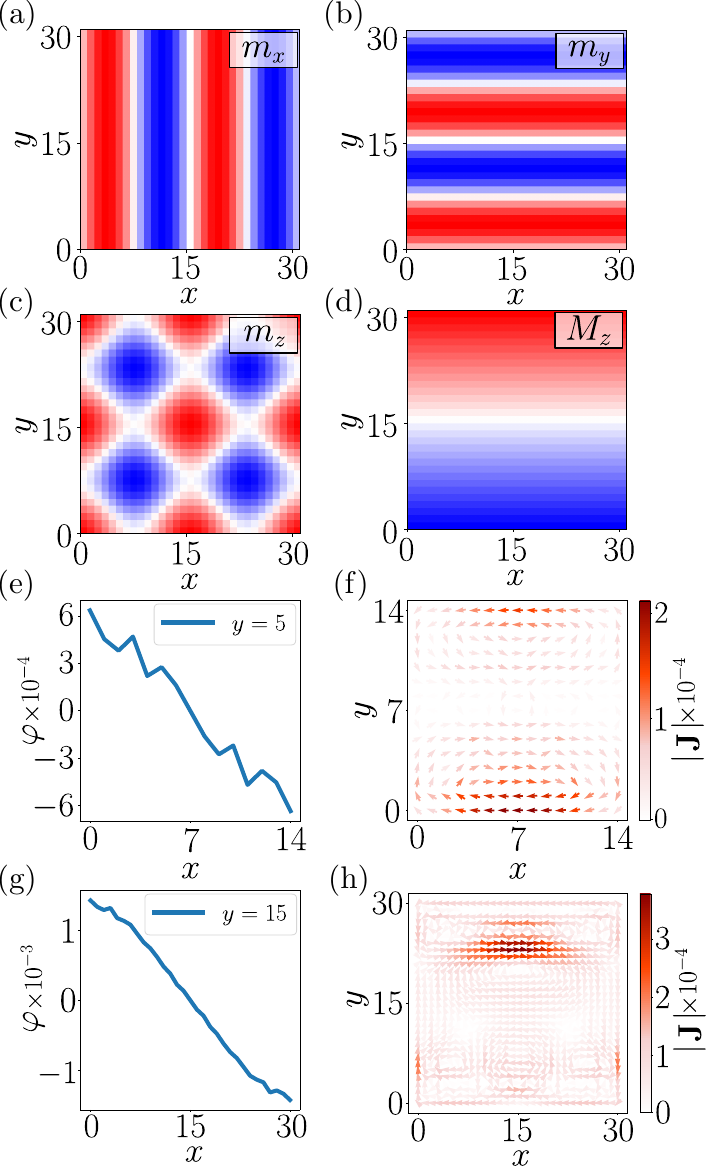}
    \caption{(a)-(c) Magnetization components for the SWC form in Eq.~\eqref{eq:SWC} with ${\cal Q} = 0.4$ and system size $31\times 31$. (d) Imposed out-of-plane magnetization gradient required for the diode effect. (e) Superconducting phase at $y=5$ as a function of position along the $x$-axis and (f) remnant total currents at each bond on a $15 \times 15$ system after requiring self-consistency, with $m = 0.05$, $2 M_z^{\textrm{max}}/\xi_{\textrm{grad}} = 0.04$, $\mu = -1.0$  and ${\cal Q} = 0.4$. (g)-(h) Same as (e)-(f) for a $31 \times 31$ system, with $2 M_z^{\textrm{max}}/\xi_{\textrm{grad}} = 0.02$ and ${\cal Q} = 0.2$. From the above, we conclude that helical superconductivity is stabilized also in the present case, with no requirement for Rashba SOC.}
    \label{fig:fig22}
    \end{center}
\end{figure}

We now proceed by numerically demonstrating the predictions for a diode effect solely due to a spatially varying magnetization profile with a zero net moment, and without any sort of intrinsic Rashba SOC, $\lambda_{\rm soc}=0$. For this purpose, we consider a  magnetization field consisting of the SWC structure described by Eq.~\eqref{eq:SWC} in addition to a uniform spatially varying out-of-plane magnetization gradient. Without of generality, we set $f=1$. Figure~\ref{fig:fig22}(a)-(c) display $2\times 2$ unit cells of the SWC magnetization components $m_x$, $m_y$ and $m_z$ entering Eq.~\eqref{eq:SWC}. The additional out-of-plane magnetization profile is shown in Fig.~\ref{fig:fig22}(d). This component is needed to obtain a helical ground state with its associated transport nonreciprocity.

In Fig.~\ref{fig:fig22}(e) and~\ref{fig:fig22}(f), we show the selfconsistent results for the induced ground state phase gradient and the absence of currents away from the edges, respectively, both characteristic of the helical phase. These results are obtained with OBC for a relatively small system size of $15\times 15$ sites. Figures~\ref{fig:fig22}(g) and~\ref{fig:fig22}(h) display the same quantities but for the case of system size $31\times 31$. As seen from Fig.~\ref{fig:fig22}(g), the superconducting phase gra\-dient $\partial \varphi /\partial x$ in the SWC-induced helical phase becomes significantly larger by doubling the system size, again featuring no currents, except close to the edge regions, see for example Fig.~\ref{fig:fig22}(h). 

Comparing the values of the phase gradients obtained from the SWC to the case with intrinsic Rashba SOC shown in Fig.~\ref{fig:fig5}, we estimate that the effective SOC ge\-ne\-ra\-ted by the SWC texture winding is approximately an order of magnitude smaller for the $31\times 31$ system. Consequently, for the relatively small system sizes available for the selfconsistent studies, given that these become very computationally demanding for such magnetization profiles, the diode effect is correspondingly reduced. 

We have verified explicitly, however, that indeed the SWC setup does feature a diode effect, and that it gets enhanced by larger systems exhibiting larger win\-ding of the SWC magnetization. We choose not to show the resulting currents here, since the small magnitude of the resulting effect is not visually discernible. The present calculations primarily serve as a proof-of-principle rather than a thorough study of the superconducting effect from magnetic textures. Further dedicated studies are required to explore the diode efficiency and further possibilities which open up from this SOC-free diode route.

\section{Discussion and Conclusions}\label{sec:discussion}

We have proposed and compared various concrete pathways to ge\-ne\-ra\-te helical superconductivity and, in turn, superconducting and Josephson diode effects. Our approach describes in a unified manner the emergence of helical superconductivity from both out-of-plane magnetization gradients and in-plane magnetic fields in Rashba superconductors. We compared the diode effect resulting from the two distinct scenarios in bulk Rashba superconductors, as well as in Josephson junctions, and found that diode efficiencies of comparable magnitude are accessible.

For the diode mechanism originating from a spatially varying out-of-plane magnetization that we propose here, we have explored uniform gradients across the Josephson junction and ferromagnetic domain walls. It is likely, however, that optimized efficiencies can be engineered by imposing other more suitable spatially varying magnetization gradients. One can also envision to use tailored magnetization gradients to guide the nonreciprocal current flow along desired paths of the 2D platform. This possibility could allow for the nonreciprocal superconducting current to be guided in circuits, without any need for lithographic etching or nanopatterning. The potential merits of such applications open the door to a number of future research directions of this sort.

Aside from out-of-plane magnetization gradients imposed on Rashba superconductors, we also showcased the Josephson diode effect in platforms where the superconducting material is SOC free. In this case, the diode effect becomes accessible due to a synthetic Rashba SOC generated by the inhomogeneous magnetization itself. However, this requires for complex magnetization profiles which are noncoplanar and wind in both spatial dimensions. As a proof-of-principle, in this work, we considered a spin-whirl-crystal magnetic texture in conjunction with the out-of-plane magnetization profile discussed earlier, and verified that both helical superconductivity and superconducting diode effects are possible in these systems. Additional dedicated studies are required to be carried out in the future, in order to identify the parameter regimes and magnetic texture profiles that can maximize the diode efficiency.

It is important to note that in this work we restricted ourselves to magnetization gradients and textures which lead to a zero net magnetization. First of all, achie\-ving zero net magnetization is advantageous for sustaining superconductivity. Second, in this work we chose to focus on such type of gradients, in order to emphasize the underlying mechanism. Considering instead other magnetic texture con\-fi\-gu\-ra\-tions, which violate this zero-sum rule, is also generally expected to give rise to a superconduc\-ting diode effect. However, in this case additional mechanisms can be at play. Interestingly, the simplest example  of a configuration which violates the zero-sum magnetization constraint can be obtained by superimposing a uniform Zeeman field on top of the out-of-plane magnetization profile discussed earlier. As it follows from Eq.~\eqref{eq:Jmag}, 
such a possibility endows the structure with enhanced control, since one ends up with an additional knob to tailor the resulting diode effect.

We also stress that the superconducting diode effect from magnetization gradients is expected in many other settings where SOC and spatially varying magnetization are at play. This may apply, for example, to the observed superconducting diode effects in twisted graphene layers in zero external magnetic field~\cite{Lin2022,Diez-Merida}. There superconductivity coexists with a spontaneously time-reversal symmetry broken state, possibly an inhomogeneous magnetic phase, known from theoretical studies to naturally appear in twisted layers with sizable interactions~\cite{breio,Banerjee2023}. In fact, one may envision using the superconducting diode effect as a probe of such spontaneously ge\-ne\-ra\-ted inhomogeneous magnetic order in quantum materials, parti\-cu\-lar\-ly in settings where transport measurements are the most appropriate probes of the system.

Finally, we point out that it would be interesting to repeat the experiments in the systems where the superconducting diode effect has been already experimentally observed, for instance in those of Refs.~\cite{Ando2020,Baumgartner2022}, but now using out-of-plane magnetization gradients. This can be possibly realized by bringing the Rashba system in pro\-xi\-mi\-ty to a ferromagnetic insulator, or, by exposing it to the fringing fields of an array of nearby nano-magnets~\cite{KotetesSuraAndersen}. In such cases, our findings predict the existence of an experimentally detectable superconducting diode effect.

\section*{Acknowledgments} 
We acknowledge useful discussions with D.~Agterberg, K.~Flensberg and M.~Geier. M.~R. acknow\-ledges support from the Novo Nordisk Foundation grant NNF20OC0060019.

\bibliography{diode}

\end{document}